\documentclass[fleqn,twoside]{article}
\usepackage{espcrc2}

\usepackage{graphicx}
\usepackage[figuresright]{rotating}

\newcommand{\AmS}{{\protect\the\textfont2
  A\kern-.1667em\lower.5ex\hbox{M}\kern-.125emS}}

\title{Local criticality close to a quantum Lifshitz point}

\author{Mucio A. Continentino\address[MCSD]{Instituto de F\'{i}sica, Universidade Federal Fluminense \\
        Campus da Praia Vermelha, Niter\'oi, 24210-340, RJ, Brazil}%
        \thanks{Tel./Fax:+55-21-2620-6735. \newline E-mail~address:~mucio@if.uff.br.
        Work supported by Brazilian Agencies, FAPERJ and CNPq }.}

\begin{document}

\begin{abstract}
Near quantum Lifshitz points the stiffness of the lifetime of the spin
fluctuations is very small. This gives rise to a regime where the critical
fluctuations are local in space but extended along the time directions. The
Fermi liquid properties in this local quantum regime is governed by a single energy
scale, the coherence temperature. These results are relevant for heavy
fermions and frustrated systems whenever there is a competition between
different instabilities in $q$-space.
\vspace{1pc}


\end{abstract}

\maketitle
The Gaussian free energy density near a quantum Lifshitz point (QLP) can be written as
\cite{mucio},
\[
\frac{f}{V}=\!\!\frac{-3T}{\pi}\!\!\int \!\!\! \frac{d^{d}q}{(2\pi)^{d}}\!\! \int \!\!\! \frac{d\lambda}%
{e^{\lambda}-1}\tan^{-1}[\frac{\lambda(T/T_{coh})}{1+q^{2}\xi%
^{2}+q^{4}\xi_{\perp}^{4}}]
\]
where the coherence temperature, $k_{B}T_{coh}=\Gamma_{L}\chi_{L}|\delta|^{\nu z}$
and $|\delta|=|J_{Q}%
-J_{Q}^{c}|$ measures the distance to a quantum critical point (QCP) (not to the QLP; this is
approached making a stiffness to vanish as discussed below). The
crossover exponent takes the value $\nu z=1$ for the problem considered here. The quantity
$\Gamma_{L}\chi_{L}$ is defined through the dynamic susceptibility of the
local moments, $\chi_{L}(\omega)=\chi_{L}(1+i\omega/\Gamma_{L})^{-1}$. Also
near the wavevector $Q$ of the incipient antiferromagnetic instability at the QCP, the
$q$-dependent interaction is written as, $J_{Q}-J_{Q+q}=Aq^{2}+Bq^{4}$.
For $T<<T_{coh}$, the free energy can be written as,
\[
f=-\frac{3V}{\pi}\frac{T^{2}}{T_{coh}}\int \!\!\! \frac{d^{d}q}{(2\pi)^{d}} \!\!\!\int \!\!\!
\frac{d\lambda}{e^{\lambda}-1}\frac{\lambda}{1+q^{2}\xi^{2}%
+q^{4}\xi_{\perp}^{4}}%
\]
which is quadratic in temperature and shows that the coherence temperature marks the onset of Fermi liquid (FL)
behavior for $T<<T_{coh}$. In the equation above, $\xi^{2}=A/|\delta|$ and $\xi_{\perp}^{4}=B/|\delta|$. In the
case the {\em stiffness} $A \neq 0$, the $q^{4}$-term is irrelevant \cite{ramaz} and we have the usual Gaussian
free energy of a nearly antiferromagnetic (AF) system in the FL regime \cite{moriya}. In three dimensions, $d=3$,
the free energy
can be integrated and we have for the specific heat in the FL regime,%
\[
C/T=\frac{4\pi^{2}}{\Gamma_{L}\chi_{L}A}\frac{V}{(2\pi)^{3}}q_{c}\left(
1-\frac{\tan^{-1}q_{c}\xi}{q_{c}\xi}\right)
\]
where $q_c \approx 1/a$, with $a$ the lattice spacing is a cut-off in momentum space. In the limit $q_{c}\xi>>1$,
for $d=3$, this yields,
\begin{equation}
C/T=6\pi^{2}Nk_{B}^{2}/Aq_{c}^{2}\label{MT}%
\end{equation}
which is the usual, non-universal, cut-off dependent result for a $d=3$,
nearly AF system.
\begin{figure}
\includegraphics[width=6.5cm]{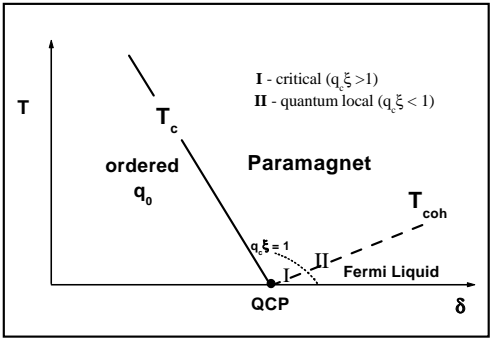}
\caption{\label{fig1}The phase diagram for a heavy fermion system.
If the stiffness \protect{$A$} is small, the critical region ($I$) is reduced and the
quantum local regime ($II$) extends very close to the magnetic quantum critical point.}
\end{figure}
We are interested here in the opposite limit, $q_{c}\xi<<1$ \cite{mucio2}. If the
stiffness  $A$ is small, as is the case close to a QLP, this
regime is relevant even close to the QCP, i.e., for $\delta$ small. We find in this limit,
\[
C/T=\frac{\pi Nk_{B}^{2}}{\Gamma_{L}\chi_{L}}\frac{1}{|J_{Q}-J_{Q}^{c}|}%
=\frac{\pi Nk_{B}}{T_{coh}}\label{LSH}%
\]
which is independent of $A$. In the same regime  the uniform susceptibility
is given by,
\[
\chi_{0}=\frac{3N\mu^{2}}{2\pi\Gamma_{L}\chi_{L}}\frac{1}{|J_{Q}-J_{Q}^{c}%
|}=\frac{3N\mu^{2}}{2\pi k_{B}T_{coh}}\label{LS}%
\]
such that, the Wilson ratio \cite{mucio},
\[
\frac{\chi_{0}/\mu^{2}}{C/\pi^{2}k_{B}^{2}T}=\frac{3}{2}=1.5\label{WR}%
\]
The electrical resistivity, for $T<<T_{coh}$ and $q_{c}\xi<<1$, is given by,
\[
\rho(T<<T_{coh})=\rho_{0}\Gamma_{L}\chi_{L}\frac{\pi^{2}}{3}(\frac{T}{T_{coh}%
})^{2}=A_{R}T^{2}%
\]
and the Kadowaki-Woods ratio, $A_{R}/({C/T})^{2}$,
\[
\frac{A_{R}}{(C/T)^{2}}=\frac{\rho_{0}\Gamma_{L}\chi_{L}}{3(Nk_{B})^{2}}%
\]
is a constant in this regime \cite{mucio}.
The results above could have been obtained, from a local free energy density,
\begin{equation}
f_{sf}^{L}=-\frac{3}{\pi}NT\int_{0}^{\infty}\frac{d\lambda}{e^{\lambda}-1}%
\tan^{-1}\left(  \frac{\lambda T}{T_{coh}}\right)
\end{equation}
which can be written in the scaling form $f_{sf}^{L}=|g|^{\nu
z}F[T/T_{coh}]$, such that, $2-\alpha=\nu z$. The hyperscaling relation,
$2-\alpha=\nu(d+z)$ \cite{mucio}, in this case is satisfied for an Euclidean dimension
$d=0$. This allows to identify the regime $q_{c}\xi<<1$ as a non-trivial
\emph{quantum local} (QL) regime where the critical correlations although being
localized in space are extended in the time directions. The effective
dimension of the critical fluctuations is consequently, $d_{eff}=z$.
Let us examine in more detail the condition $q_{c}\xi<<1$ or $q_{c}%
\sqrt{A/|\delta|}<<1$. This implies that, for example, in the $T-|\delta|$ plane there is a crossover line in the
non-critical side of the phase diagram, $q_{c}\xi(T)=1$ which separates two regimes: the true critical regime
where $q_{c}\xi(T)>>1$ and the quantum local, for $q_{c}\xi(T)<<1$, where correlations are mainly in the time
directions. It is interesting that as the stiffness $A \rightarrow 0$, the critical region shrinks and the quantum
local regime holds closer and closer to the quantum critical point. The Fermi liquid energy scale in this regime
is the coherence temperature, {\em which is independent of the stiffness} $A$ and the characteristic time scale
is, $\hbar \tau_{\xi}^{-1}=k_B T_{coh}$. Eventually, when $A=0$, the QLP is reached and the critical behavior is
governed by the associated fixed point \cite{ramaz}. On the other side of the QLP, for $A<0$, the function
$f(q)=Aq^{2}+Bq^{4}$, vanishes for $q_{0}=\pm\sqrt {|A|/2B}$. Near $q_{0}$, we have,
$f(q)=-A^{2}/4B+(2|A|)(q-q_{0})^{2}$. The first term is incorporated in $\delta_{h}=\delta+A^{2}/4B$, such that,
the correlation length now diverges at the \emph{helicoidal} transition, $|\delta_{h}|=0$ \cite{ramaz}. The QLP
governs the critical behavior only for $A=0$. $A$ is a relevant perturbation and for any $|A| \ne 0$ the quantum
critical behavior is controlled by the usual Hertz fixed point associated with spin-density wave transitions
\cite{hertz}. The zero temperature transition from the paramagnetic to the helicoidal metal is in the same
universality class of the para-antiferromagnetic metal transition ($z=2$). For $d=3$ they are above the critical
dimension and are described by a Gaussian theory \cite{mucio}.
\begin{figure}
\includegraphics[width=7.5cm]{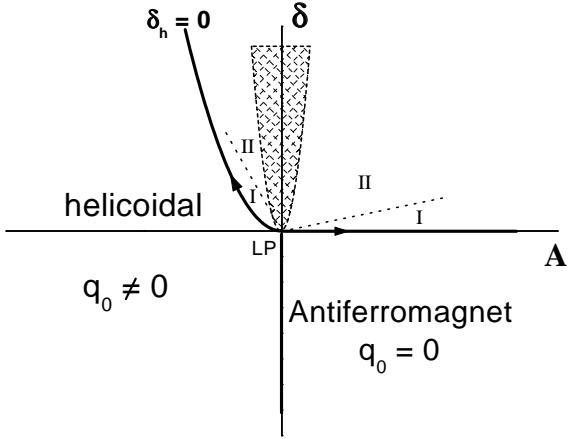}
\caption{\label{fig2} Phase diagram close to a quantum Lifshitz point \protect{\cite{ramaz}}.
In 3d, the lines of para-AF ($|\delta|$=0) and para-helicoidal ($|\delta_h|$=0) phase transitions are governed by 
quantum critical
points with Gaussian exponents and $z=2$. Regions $I$ and $II$ have the same meaning as in Fig.1. The dashed region is 
that of Lifshitz quantum criticality  \protect{\cite{ramaz}}. Arrows represent the flow of the renormalization group 
equations. }
\end{figure}
In the nearly helicoidal metal the correlation length,
$\xi_{h}^{2}=2|A|/|\delta_{h}|$. There is also in this case a crossover line
$q_{c}\xi_{h}(T)=1$, such that, for  $q_{c}\xi_{h}(T)<1$ a quantum local
regime exits where the system presents a similar behavior as in the nearly AF metal.

There are both direct \cite{aeppli,lacroix} and indirect evidence that heavy fermions and frustrated magnets are
close to quantum Lifshitz points. The most dramatic is the heaviness of the quasi-particles which indicates the
smallness of the stiffness $A$ (Eq.\ref{MT}). How quantum local behavior manifests along the usual non-Fermi
liquid path, $|\delta|=0$, $T \rightarrow 0$ and the role of dangerously irrelevant variables are the subject of
present investigations.

We have shown the existence of a significant region in the phase diagram of systems close to a QLP where the
physics is governed by fluctuations local in space but extended in time.

\end{document}